\documentstyle[12pt,epsfig,rotate]{article}

\newcommand{\afh}{\mbox{$ Ar/CF_4/CH_4 $}}
\newcommand{\afo}{\mbox{$ Ar/CF_4/CO_2 $}}

\begin{document}

\begin{titlepage}
 
\pagebreak

\vspace*{0.1cm}
\begin{center}
{\huge Institute of Theoretical }
\end{center} 
\begin{center}
{\huge and Experimental Physics }
\end{center} 
\begin{center}
{\huge Preprint  15-01 }
\end{center} 
\vspace{3.0cm}
\begin{center}
{\large M.Danilov, Yu. Gilitsky, T. Kvaratschellia, L.Laptin, I.Tichomirov,}
\end{center} 
\begin{center}
\vspace{0.05cm}
{\large M.Titov, Yu.Zaitsev}
\end{center} 

\begin{center}
\vspace{2.cm}
\end{center}
\begin{center}
\vspace{0.1cm} 
{\large\bf Aging Studies of Large Area Proportional }
\end{center}
\begin{center}
\vspace{0.1cm} 
{\large\bf Chambers under High-Rate Irradiation }
\end{center}
\begin{center}
\vspace{0.1cm} 
{\large\bf with $CF_4$-based Mixtures (PART 1). }
\end{center}
\begin{center}
\vspace{0.1cm} 
\end{center}
 
\vspace{1.5cm}
\begin{center}
{\large \bf  PART 1 }
\end{center}

\vspace{5.5cm}

\begin{center}
{\large\bf Moscow 2001}
\end{center} 

\end{titlepage}

\newpage

\vspace*{7.5cm}

\large 

 Experimental conditions at the HERA-B experiment impose very strong 
requirements for gaseous detectors.
The charged particle fluxes through the HERA-B tracking system,
varying with the radial distance $R$ from the beam line,
are about $2 \times 10^{7}/R^{2}$ particles per second,
and comparable to those that will be encountered by LHC experiments.

 The severe radiation environment 
of the HERA-B experiment leads to a maximum
charge deposit on a wire, within the muon detector, 
of 200 mC/cm per year. We report recent results of aging studies 
performed by irradiating proportional wire chambers
filled with $Ar/CF_4/CH_4$ (74:20:6), $Ar/CF_4/CH_4$ (67:30:3),
$Ar/CF_4/CO_2$ (65:30:5), $Ar/CF_4$ (70:30), $CF_4/CH_4$ (90:10),
$CF_4/CH_4$ (80:20) mixtures in a three different  experimental setups.
 The size of the irradiation zone varied in the tests 
from 1~cm  up to 500~cm. 
Our experience shows that the aging rate depends not only on the total
collected charge, but, in addition, on the mode of operation and area
of irradiation.
 The possible 
application of these results to the construction of a large area 
gaseous detectors for operation in high rate environments is presented.

\large
\clearpage
\pagenumbering{arabic}
\setcounter{page}{1}

\vspace*{2.2cm}

\section{Introduction}

 The HERA-B experiment is a hadronic B-factory at DESY, where 
B-mesons are produced via interactions of 920~GeV beam halo
protons with an internal target~\cite{Proposal,TDR}. 
In order to reach the necessary 
production rate of $b$-quarks, the HERA-B detector must operate at an
extremely high interaction rate of $\sim$~40~MHz, corresponding 
to an average of 4 superimposed interactions per HERA proton bunch 
crossing. Such a high rate of inelastic interactions results 
in events with large charge and neutral multiplicities -  up to
several hundred particles every 96 ns - and a harsh radiation
environment for the constituent detectors. 
The peak integrated doses in the HERA-B tracking system
are about 
1~Mrad/year for the Inner Tracker and  500~$\frac{mC}{cm~wire}$ 
per year for the Outer Tracker.

 The high irradiation levels in HERA-B led to several painful 
experiences with aging problems in gaseous detectors. 
 Although irradiation of small area chambers with $X$-rays showed 
that the original Inner and Outer Tracker prototype 
chambers are able to 
tolerate very large X-rays doses, corresponding to several years 
of operation without visible aging effects, in the HERA-B
environment both chambers died within hours. In a
subsequent extended R$\&$D program these problems were solved,
either by going to a revised design of the 
Inner Tracker (e.g. introducing GEM foils in the original 
diamond coated MSGC)~\cite{sauli,schmidt,zeuner} 
or by changing the original concept for the drift tubes of the Outer 
Tracker (surface conductivity, gas mixture, production 
materials)~\cite{stegmann,hohlmann,kolanoski,dissert1}.
  
 Strong restrictions on the gas choice are  
imposed by the operating conditions in the muon detector
(the maximum accumulated 
charge on the wire can reach 200~$\frac{mC}{cm~wire}$ per year).  
 In a previous paper~\cite{max},
studies of the aging performance of muon proportional 
chambers under sustained irradiation 
with $Fe^{55}$ and $Ru^{106}$ sources and in a 100 MeV $\alpha$-beam
were presented.
  Despite the negligible gas gain loss measured after the long-term 
irradiation with radioactive source, rapid aging effects 
during irradiation with the 100~MeV $\alpha$-particles 
were observed in an $Ar/CF_4/CH_4$ (74:20:6) mixture.
 From these results it is evident that from the accumulated charge 
alone, it is difficult to combine the data from the different 
radiation sources into one consistent model.
 The difference between these results could be 
partially attributed 
to the type of ionization particles, 
the size of the irradiation area, gas gain, gas flow rate 
or some other variations in the operating parameters.
 As long as the data from different aging setups can not be combined
properly, it is very difficult to give an extrapolation about the 
lifetime of the real detector.
Therefore, we have tested the aging performance of 
muon proportional chambers filled with
$\afh$ (74:20:6), $\afh$ (67:30:3) and $\afo$ (65:30:5),
$Ar/CF_4$(70:30), $CF_4/CH_4$(90:10), $CF_4/CH_4$(80:20)
mixtures in the HERA~-~B environment, under conditions as close as 
possible to the real ones.  The results of these aging tests 
are summarized in this paper.

\section{ Aging Studies in a HERA-B Environment}

\subsection{ Experimental Setup }

 Three different types of gas proportional chambers make up the
muon system: tube, pad and pixel~\cite{TDR}. 
In the following we report on aging studies performed with
aluminum proportional tube chambers in the real HERA-B environment.

  The main objective of these tests was to prove that 
muon chambers are capable of operating after exposure to an
integrated charge of 600~$\frac{mC}{cm~wire}$,
which is expected after three years of operation in the hottest 
region, without loss in performance.



 The tube chamber is a closed-cell proportional wire chamber 
made from an aluminum profile~(wall thickness 2~mm)
with a drift cell 14~$\times$~12~mm$^2$ in cross section.
Fig.~\ref{tube1}  shows a schematic drawing of a tube chamber.
A gold-plated tungsten wire of 45~$\mu$m diameter and a length 
of nearly 3~m is stretched inside each cell and fixed 
mechanically with pins at the chamber endcaps.

 In order to test the aging properties under conditions as
close as possible to real ones, three tube chambers with
16 drift cells each and a cross section exactly as 
for the production version, but of a shorter length of
50~cm, were placed between the electromagnetic calorimeter
and the muon absorber in the HERA-B experiment, where the charged
and neutral particle fluxes are extremely high~\cite{TDR}. 
 Initially, two of the chambers filled with
$Ar/CF_4/CH_4$ (67:30:3) and $Ar/CF_4/CO_2$ (65:30:5) mixtures
were subdivided to allow groups of anode wires (zones)
to be operated at five different high voltages. 
Fig.~\ref{tabl1} shows a simple schematic representation of the
experimental setup, numbering
scheme for the wires, and summarizes the operating conditions.
 The applied voltages were chosen in a way
to obtain comparable accumulated charges in zones 2,3,4,5.
 Between any two zones there was a reference wire for which
high voltage was not applied.
The total collected charge in each HV zone 
was determined by integrating the continuously recorded anode current.
Due to variations in the rate of $pN$-interactions
in the HERA-B experiment, 
the radiation intensity and thus the average current
varied during the studies by a factor of 4, thus
excluding a possibility to investigate the aging behavior 
as a function of rate.
 It should also be noted that tests 
with $Ar/CF_4/CH_4$ and $Ar/CF_4/CO_2$ mixtures 
have been done at about
one order of magnitude higher radiation levels than expected
in the muon detector (20~nA/cm),
thus allowing a study of the aging properties of the gases
in a more severe radiation environment.
The third chamber was subdivided into 4 zones, 
each containing two or three wires, as shown in Fig.~\ref{tabl1}.
  Radiation studies with mixtures
$CF_4/CH_4$ (90:10), $CF_4/CH_4$ (80:20), $Ar/CF_4$ (70:30) 
were performed on different chamber wires in order
to investigate the influence of materials and possible 
trace contaminants in the gas system, as well as water addition and 
cathode cleanliness, on the aging performance.

 For each chamber a pre-mixed gas was transported 
by a 150~m stainless steel tube followed by a polyamid tube 
connected directly to the chamber inlet. 
The addition of water to the gas was done indirectly, 
by varying the  length of the polyamid tube,
which is transparent to $H_2O$, for a fixed flow rate.
The gas flow within
the chamber was serial: the gas inlet was connected to chamber
cell N1, and the gas outlet to cell N16.
The gas outlet of the chamber was connected via a 50~m 
stainless steel tube to the input of a gas chromatograph, 
which was used to analyze the concentration of species
($Ar$, $CF_4$, $CH_4$, $CO_2$, $N_2$, $O_2$, $H_2O$)
in the effluent gas stream.
Usually, measured concentrations of $N_2$ and $O_2$ 
were at the level of 150-250 ppm.


 Periodically (typically once/month) chamber characteristics
were studied in a 3~GeV electron beam with an intensity 
$\sim$~1~$kHz/cm^2$. A reduction in efficiency measured with 
electrons
at different positions along the wire was used as the main
information about the loss in performance. 
In addition, during irradiation in the HERA-B environment
several other types of performance degradation
with distinct symptoms also indicate the
appearance of aging effects: dark current, special
'switch-on' current behavior (decrease
in the anode current after high voltage is applied, since
a certain time is necessary to establish the equilibrium
charge density on the aged wire surface).
 This phenomena has been widely described in the 
literature~\cite{algeri, tsarnas,fraga}.
And finally, an interesting phenomena was observed 
for all aged wires operated with an $Ar/CF_4/CH_4$ mixture:
the operating current was dependent upon the gas flow rate
(an increase of gas flow led to
an increase in the anode current).
 We will come back to the experimental study 
of this specific aging phenomena in a subsequent paper~\cite{max1}.
At the same time, for all wires without gain reduction,
neither 'switch-on' current behavior nor 
dependence of operating current upon the flow rate have
been observed.
 Thus, the appearance of any type of performance degradation
for a particular group of anode wires could also serve 
as a manifestation of aging effects.

 Before the start of the aging run, tube chamber efficiency 
and singles counting rate as a function of high voltage  were 
measured for different gas mixtures in the electron beam
(see Fig.~\ref{fig3}a,b).
The readout of the chambers is based
on the ASD-8 amplifier shaper discriminator 
chip~\cite{electronics}. Therefore,
real afterpulses and multiple hits coming from ASD-8
per signal both contribute to the rise in singles rate curves,
above the nominal beam intensity, with increasing high voltage.
 According to results reported in~\cite{max2},
operating voltages for the tube chamber were chosen for the following: 
$Ar/CF_4/CH_4$ (74:20:6)~-~2.25~kV,  
$Ar/CF_4/CH_4$ (67:30:3)~-~2.45~kV, 
$Ar/CF_4/CO_2$ (65:30:5)~-~2.4~kV,
corresponding to the baseline
gas amplification  $\sim~3 \cdot 10^4$.

\subsection{ Aging in an $\afh$(67:30:3) gas mixture}

 Radiation tests with an $\afh$ (67:30:3) mixture, which is 
similar to $Ar/CF_4/CH_4$ (74:20:6), have been carried out
in the real HERA-B environment.
 Due to the reduced $CH_4$ content in $\afh$ (67:30:3), better aging
properties of this mixture were expected. This was also confirmed 
during irradiation with a $Ru^{106}$ source, where stable
operation up to a total collected 
charge of 2~$\frac{C}{cm~wire}$ was achieved~\cite{max}. 
Since water added to the mixture in small concentrations
is believed to prevent polymerization of hydrocarbons,  
we used 500~$ppm$ of $H_2O$ from the beginning of the tests.
 Typically, a chamber was flushed at a flow rate of 3~liters/hour,
corresponding to two volume exchanges per hour.

 First results in the HERA-B environment showed that  
aging effects
in the $\afh$(67:30:3) + 500 $ppm$ of $H_2O$ mixture depend on 
the gas amplification (high voltage) and the size of the
irradiation area.



  After 10 days of exposure which resulted in a collected
charge of $\sim$~25~$\frac{mC}{cm \cdot wire}$ 
electron beam tests showed
gain reduction for two (N14,N15) from three 
wires operated at high voltage 2.65~kV.
 After an accumulated charge of $\sim$~100~$\frac{mC}{cm \cdot wire}$
all three wires (N13,N14,N15) in zone 5 were already inefficient.
 A similar aging behavior has been observed for the wires 
irradiated at 2.6 kV, however their lifetime 
was slightly longer. After an 
accumulated charge of $\sim$~80~$\frac{mC}{cm \cdot wire}$, 
gain reduction
has been found for one wire N11 and approximately at a dose of 
170~$\frac{mC}{cm \cdot wire}$
electron beam tests revealed efficiency loss
in both cells (N10,N11). 
In addition to a gain reduction for the wires irradiated
at 2.6~kV and 2.65~kV, operation of these cells
in the HERA-B environment was accompanied 
by the 'switch-on' current behavior and the appearance of the 
dependence of the operating current on the gas flow rate.
Dark current (or Malter effect) in these zones also started 
to appear and could lead to a subsequent detector degradation.

It has to be noted that the
efficiency loss for wires operated at 2.6~kV and 2.65~kV 
was more severe in the direction of the serial gas flow.
 Initially aging effects in zones 4 and 5 appeared 
on the last wires (N11,N15) in the serial gas flow, although
the radiation intensity for these wires was the lowest in each zone.
 Fig.~\ref{graf1} shows an example of how the efficiency profile 
developed in the direction of gas flow for wire N11 with 
the accumulated charge. 
The loss in performance for cell N11
with increasing 'usage' of the gas is obvious.
 Wires (N13,N14,N15) were irradiated up to a
dose of $\sim$~200~$\frac{mC}{cm \cdot wire}$, 
while aging tests in zones 2,3,4 were continued
up to collected charge $\sim$~400~$\frac{mC}{cm \cdot wire}$. 
Due to the smaller current
density the total accumulated charge  in zone 1
was only $\sim$~100~$\frac{mC}{cm \cdot wire}$.
 We did not observe any onset of dark current in cells N1-N8
during the whole period of irradiation.
 After the aging run, the chamber performance was studied
in an electron beam. All wires operated at 2.6~kV, 2.65~kV 
showed  strong anode aging.
 At the same time, scanning along the wires N5,N7,N8
with an electron beam  revealed
several distinct spots with a typical size of several $cm$,
where local inefficiencies were detected 
(measured efficiency at 2.4~kV was $\sim$~90-98~$\%$).
 No change in performance was found for wires N1,N2~(2.25~kV),
N4~(2.5~kV), and all reference wires.
 

In addition to measurements in the electron beam,
at the conclusion of the run wires irradiated 
at 2.6~kV and 2.65~kV were taken for surface analysis
at `Digital Analytical Scanning Microscope JSM-6400'.
 The scanning electron microscope (SEM) yields information on
the morphology of the wire surface by imaging with scattered 
and secondary electrons, and 
energy dispersive X-ray spectroscopy (EDX) analysis gives 
the atomic composition of the surface material.
Usually, micrographs of anode wires were taken at 
SEM accelerating voltage 20~kV (sometimes -~5~kV),
while cathodes were analyzed at 5~kV.
 The increased sensitivity to surface deposits at 5~kV is due
to the reduced penetration of the lower energy primary electrons.
Fig.~\ref{sum1} shows SEM micrographs and EDX analysis of 
wire N13 specimen taken near the gas inlet 
and wire N15 specimen  taken near the gas outlet
(150~cm downstream from the beginning of zone 5 
in the serial gas flow).
 EDX analysis of these specimens confirmed the presence
of polymers on the wires, containing carbon and
fluorine as the only detectable elements (SEM is insensitive   
to hydrogen) (see Fig.~\ref{sum1}). 
Moreover, analysis of the wires (N13,N14,N15) 
have shown that the $F/C$ ratio in the surface deposit
increases in the direction of the serial gas flow.
 SEM micrograph of wire N11 irradiated at 2.6~kV
was also taken and is shown in Fig.~\ref{gr53}.
In the whisker emission 'EDX' spectrum there were 
identified lines
corresponding only to $C$ and $F$ elements.

 
At this point we decided to study the influence of water
on aging performance and continued tests with an increased
$H_2O$ content (1400~$ppm$), while using
the previously irradiated  wires in zones 1,2,3.
 During three monthes of operation, 
approximately 400~$\frac{mC}{cm \cdot wire}$ 
were accumulated for wires irradiated at 2.5~kV and  2.55~kV
and $\sim$~100~$\frac{mC}{cm \cdot wire}$ for wires at 2.25~kV 
in addition to those collected in the previous run, which resulted
in a total radiation dose of $\sim$~800~$\frac{mC}{cm \cdot wire}$ 
in zones 2,3 and $\sim$~200~$\frac{mC}{cm \cdot wire}$ in zone 1. 
 Further scanning along the wires with an electron beam have shown that
small inefficient spots on wires N5,N7,N8 had vanished, and
all wires became fully efficient again. 
No change in performance was also observed for wires N1,N4.
At the same time, an efficiency drop was detected for 
one of the wires, N2, operated at 2.25~kV, 
being more severe near the chamber endcaps 
(measured efficiency at 2.4~kV was $\sim$~85~$\%$)
than in the center of the wire ($\sim$~97~$\%$),
where the radiation intensity was the largest. 
 Upon completion of the beam studies, all wires from the 
chamber were analyzed under the electron microscope.
EDX analysis of irradiated wires N1,N4,N5,N7,N8
and all reference wires revealed only $Au$ peaks thus
confirming the absence of the polymer coating. 
Fig.~\ref{ttt111} shows the SEM micrograph and EDX spectrum for wire N5.
 However, SEM imaging of wire N2 revealed 
a surface deposit consisting 
of traces of $Si$ and $O$ (see  Fig.~\ref{ttt111}). 
The absence of $Si$-containing deposits on all other aged wires 
operated with $Ar/CF_4/CH_4$ indicates that $Si$-polymerization 
was not a dominant aging process in our tests, 
and particularly for cell N2 
could be a result of contamination from chamber endcaps. 
 Moreover, gaseous discharges in an $Ar/CF_4/CH_4$ mixture
could provide additional resistance for $Si$-deposits,
which react with fluorine radicals to form volatile $Si F_4$.
In plasma polymerization, discharges of $CF_4/H_2$ mixtures are 
successfully used for $Si O_2$ etching, while $CF_4/O_2$ plasmas
selectively etches $Si$~\cite{kushner}.



 In order to study the influence of water on aging
performance in $Ar/CF_4/CH_4$,  
we have carried out new radiation tests. 
For this all chamber cells were restrung with new wires.
 With an electron beam rate of 1~$kHz/cm^2$,
no difference in efficiency
between previously irradiated and reference cells was found.
However, the dark current observed 
in the previous run from wires  N10,N11,N13,N14,N15
indicates that 
surface deposits could be present on the cathodes in these cells.
 Although a subdivision of anode wires into 4 high
voltages groups was slightly changed for this run:
zone~1 (wires N1,N2) - 2.25~kV, zone~2 (wires N4,N5,N6) - 2.5~kV,
zone~3 (wires N8,N9) - 2.6~kV, zone~4 (wires N14,N15,N16) - 2.65~kV,
this did not affect significantly the average current density in each zone.
Initially, radiation tests with the 
previously used mixture $Ar/CF_4/CH_4$ (67:30:3)
+ 1400 $ppm$ of $H_2O$ were performed up to radiation doses:
40~$\frac{mC}{cm \cdot wire}$ in zone 1, 
80~$\frac{mC}{cm \cdot wire}$ in zone 2,
100~$\frac{mC}{cm \cdot wire}$ in zone 3,
120~$\frac{mC}{cm \cdot wire}$ in zone 4.
 During the whole period of irradiation, 
we observed dark current for the wires operated at 2.65~kV.
At this point all polyamid tubes were exchanged 
with  stainless-steel lines, thus excluding the presence of water
in the chamber. 
 Several types of performance degradation (dark current,
steadily decreasing anode current) started to appear 
for wires operated at 2.5~kV, 2.6~kV, 2.65~kV
at a level of an accumulated charge 
$\sim$~15-30~$\frac{mC}{cm \cdot wire}$.
 These results were especially surprising for wires
irradiated at 2.5~kV, since no
gain reduction has been observed 
up to 800~$\frac{mC}{cm \cdot wire}$
in $Ar/CF_4/CH_4$ (67:30:3) mixture 
with 500 $ppm$ and 1400 $ppm$  of $H_2O$.
At the end of the aging run without water, 
the following radiation doses were collected:
20~$\frac{mC}{cm \cdot wire}$ in zone 1,
70~$\frac{mC}{cm \cdot wire}$ in zone 2,
85~$\frac{mC}{cm \cdot wire}$ in zone 3,
65~$\frac{mC}{cm \cdot wire}$ in zone 4. 
(High voltage in zone 4 was switched off earlier than in other zones).
  It might be noted that in sharp contrast to wires in other zones 
no performance degradation for wires in zone 1 was observed 
up to the end of the run.


 Studies of chamber efficiency in an electron beam
confirmed significant gain reduction for all wires operated at
2.5~kV, 2.6~kV, 2.65~kV. Moreover, a
strongly nonconductive polymer coating present on the wires 
complicated the measurement of efficiency due to
the time-dependent amplitude suppression resulting from
charging-up of aged wire surfaces~\cite{algeri}. 
When the chamber was opened for inspection,
 dark whiskers randomly distributed on all 
irradiated wires in zones 2,3,4 were observed.
  Fig.~\ref{gr52} presents typical micrographs of the 
shaped growth deposits revealed by SEM on the wires N5 (2.5~kV)
and N9 (2.6~kV), which are very similar to those observed 
after irradiation with $Ar/CF_4/CH_4$ (67:30:3) + 500 $ppm$ $H_2O$.
 The structure of these 'whiskers' suggests that they are caused
by debris attaching to the anode wire and could serve as nucleation
points for further deposits.
 It is also important to note that
the deposit thickness was clearly irregular,
with some areas appearing relatively clean while others had 
'whisker' type deposits. 
 Moreover, even in the 'whiskers' free region, surface deposits
consisting of $C$ and $F$ elements and masking the underlying
gold signal, were identified (see Fig.~\ref{gr52}).
 Scanning along wires N1,N2 irradiated at 2.25~kV,
revealed a loss in efficiency for wire N1 with a maximum
drop in the center of the wire
(measured efficiency at 2.4~kV was $\sim$~93~$\%$),
where the radiation intensity was the largest.
No change in performance was observed for wire N2.
In contrast to results in other zones, these data demonstrate
a dependence of the degree of aging on the irradiation intensity,
and thus the accumulated charge, 
and not on the direction of the serial gas flow.
 SEM analysis at 20~kV of wire N1 
revealed only gold, while the EDX spectrum  at 5~kV 
is dominated by an intense $C$ peak masking the $Au$ signal.
 The different texture of the wire deposits at 5~kV and 20~kV, 
as a result of increased SEM sensitivity at
5~kV to the thin carbon layer, is clearly evident from Fig.~\ref{pic5}.
It is also notable
that the deposits on the wire N1 
are carbonaceous, without incorporation of $F$ into the
polymer structure.
This is in contrast to typical deposits detected after 
irradiation with higher voltages.

Unfortunately, the relatively small accumulated charge in zone 1,
leaves us unable to conclude whether the
difference in chemical composition of polymers
in zone 1 and 2-4
can be attributed to the change in the polymerization mechanism 
as a result of increased plasma density 
in the avalanche for higher voltage, or simply
indicates that the polymerization in an
$Ar/CF_4/CH_4$ (67:30:3) mixture starts from the 
deposition of a carbon layer on the anode wire.
 Detailed discussion of the results presented here
will be reported in a companion paper~\cite{max1}.

 Several cathode pieces were also analyzed under the electron
microscope. As can be seen from Fig.~\ref{pic6}a,  
a thin deposit, consisting of $C$, $F$ and $O$,
was detected on the irradiated $Al$ surfaces.
 Fig.~\ref{pic6} also shows that the maximum $F$ abundance was found
on cathode cells N5 and N8, operated
up to a dose $\sim$~1000~$\frac{mC}{cm \cdot wire}$
in an $Ar/CF_4/CH_4$ (67:30:3) mixture with 500 $ppm$ of $H_2O$,
1400 $ppm$ of $H_2O$, and without water.
Analysis of cathode cell N10, after
irradiation with $Ar/CF_4/CH_4$ (67:30:3) + 500 $ppm$ of $H_2O$
up to 400~$\frac{mC}{cm \cdot wire}$ and cathode cell N6,
after irradiation with $Ar/CF_4/CH_4$ (67:30:3) + 1400 $ppm$ of $H_2O$
up to 150~$\frac{mC}{cm \cdot wire}$,
revealed a comparable $F$ abundance.
 Since the operating conditions for wires N5 and N6 were nearly
identical during the last run, the difference in
 the $F$ abundance in these cells could be 
explained by the fact that polymer deposition 
on the $Al$ surface in cell N5 also occured 
in the previous run with water, when 800~$\frac{mC}{cm \cdot wire}$
were collected.
 Similar $F$ abundances on cathode cells N6 
and N10, after different accumulated charges,
indicates a larger rate of fluorocarbon deposition 
on $Al$ surfaces without water.
 However,  it is not clear
whether polymers deposited on the cathodes in the previous run
(with 500~$ppm$ and 1400~$ppm$ of $H_2O$)  could provoke
extremely rapid aging effects during operation without water,
or if such an aging rate is defined solely by the kinetics 
of $Ar/CF_4/CH_4$ wire avalanches without water,
under particular discharge conditions. 
 
{\bf {Our results clearly demonstrate that the addition of water to the
mixture $Ar/CF_4/CH_4$ (67:30:3) substantially increases the
lifetime of the
aluminum proportional chambers and decreases
the rate of polymer deposition on the anode wires and cathode surfaces.
Although under some conditions (gas gain $< 10^5$, water addition 
$>1000~ppm$, gas flow 3~l/h) tube chambers 
can be operated up to a nominal lifetime (600~$\frac{mC}{cm \cdot wire}$),
the strong dependence of aging properties on high voltage
and size of irradiation area completely ruled out the
$Ar/CF_4/CH_4$ (67:30:3)
mixture as a candidate for operation in the muon chambers
at the HERA-B experiment.}}

\subsection{ Aging in an $\afo$(65:30:5) gas mixture}

 Our intensive aging studies have shown that both
$Ar/CF_4/CH_4$ (74:20:6) and $Ar/CF_4/CH_4$ (67:30:3)
mixtures failed to fulfill radiation hardness requirement and 
can not be used for operation in the muon detector.
 In order to test the suitability of the $Ar/CF_4/CO_2$ (65:30:5) +
1000~$ppm$ of $H_2O$ mixture,
radiation studies were performed at five different high voltages
(gas gain ranges from $10^4$ to $ 3 \cdot 10^5$). 
 We have employed a gas flow rate of 3~l/hour, 
corresponding to 2~volume exchanges per hour.




 No change in performance of aluminum proportional chambers 
operated with $Ar/CF_4/CO_2$ (65:30:5) + 1000~$ppm$ mixture
has been observed during the whole run up to collected charge 
$\sim$~700~$\frac{mC}{cm \cdot wire}$ in zones~2-5.
(The accumulated charge for the wires in zone~1 was 
$\sim$~170~$\frac{mC}{cm \cdot wire}$).
Fig.~\ref{graf41} shows efficiency profiles along wires N5,N10,N13
after an accumulated charge of 700~$\frac{mC}{cm \cdot wire}$
in the $Ar/CF_4/CO_2$ (65:30:5) + 1000~$ppm$ mixture.

At the conclusion of these tests, the
 tube chamber was opened for inspection.
 Fig.~\ref{pic11} shows typical SEM micrographs 
of the wires irradiated at high voltages:
2.5~kV (wire N5), 2.6~kV (wire N10), 2.65~kV (wire N13).
 Although SEM imaging revealed some local 
deterioration of the smooth wire 
surface after a long-term operation at high gain 
$\sim 3 \times 10^{5}$ (see Fig.~\ref{pic11}), the elemental composition
of this wire is dominated by $Au$ peak (at  20~kV SEM voltage)
and there is no evidence of other chemical elements.
The appearance of a few wire specimens was slightly more black
than that of a new one, however
detailed SEM investigation of all irradiated wires 
have shown apparently clean 
surfaces with only a few point-like deposits ($Si$,$Al$) 
with typical size of 1-2~$\mu m$, that
did not result in local inefficiencies.

 While SEM demonstrated the absence of polymers on the  anode 
wires, EDX analysis revealed trace amounts of $C$, $O$ and $F$
on the $Al$ cathodes in cells N13,N14,N15, irradiated
at 2.65~kV (see Fig.~\ref{pic22}).  
 For cells N7 and N8, operated with a lower voltage of 2.55~kV, 
the relative
abundances of $C$, $O$, $F$ on $Al$ cathodes were greatly
reduced. Cathodes in cells N4 and N5, operated at 2.5~kV, 
were almost indistinguishable from the reference ones.
 Such fluorocarbon films on the  cathode surfaces
did not lead either to dark current 
or to efficiency loss.
 The appearance of fluorocarbon deposits on irradiated cathodes 
in $CF_4$-containing mixtures has been reported in the literature. 
 As was shown in~\cite{chemmod}, 
trace fluorocarbon deposits formed on cathodes 
after irradiation in pure $CF_4$
resulted in a loss of gas gain, rather than in a self-sustaining 
discharge.
Although in other tests an increased layer
of fluorine and oxygen was observed on the cathode,
no change in gas gain was observed during long-term 
irradiation~\cite{dissert,denisov}.

{\bf { Using $Ar/CF_4/CO_2$(65:30:5) + 1000 $ppm$ $H_2O$ mixture,
no loss in performance of aluminum proportional chambers
was observed up to  collected
charge $\sim$~700~$\frac{mC}{cm \cdot wire}$, 
whilst irradiating with different gas gains.
 Analysis of cathode surfaces revealed  
trace amounts of $C$, $O$ and $F$
for tubes operated at high gas gain $\sim 10^5$, 
while for a baseline gain of $3 \cdot 10^4$,
exposed cathodes were almost indistinguishable from 
reference ones. 
 Therefore, $Ar/CF_4/CO_2$(65:30:5) + 1000 $ppm$ $H_2O$
 mixture is currently used for the muon chamber operation.}}

\newpage

\begin{figure}
\caption{ Schematic drawing of the single layer tube chamber. 
Dimensions are  indicated  in units of mm.    
\label{tube1} }
\end{figure}

\begin{figure}
\caption{ Schematic representation of the experimental system 
and numbering scheme for wires in the chamber.
Operating conditions (high voltage 
and average current density) in each zone are also indicated.
\label{tabl1}}
\end{figure}

\begin{figure}
\caption{ a) Tube chamber efficiency 
as a function of high voltage; b) Singles counting rate as a 
function of high voltage.
  \label{fig3}}
\end{figure}

\begin{figure}
\caption{ Efficiency profiles along wire N11 for three
 different accumulated charges: 80~$\frac{mC}{cm \cdot wire}$,
 170~$\frac{mC}{cm \cdot wire}$, 270~$\frac{mC}{cm \cdot wire}$,
 measured in an electron beam at 2.4~kV with $Ar/CF_4/CH_4$ 
 (67:30:3) mixture. These results are to be compared with 
 the efficiency of reference wires $> 99~\%$ at 2.4~kV.
\label{graf1}}
\end{figure}

\begin{figure}
\caption{ SEM micrographs and EDX analysis of  
wire N13 specimen taken near the gas inlet 
and wire N15 specimen taken near the gas outlet
(150~cm downstream from the beginning of zone 5
in the serial gas flow).
   \label{sum1}}
\end{figure}

\begin{figure}
 \caption{ SEM micrograph of wire N11 irradiated at 2.6~kV.
After a dose of 400~$\frac{mC}{cm \cdot wire}$ in an $Ar/CF_4/CH_4$
(67:30:3) + 500~$ppm$ $H_2O$ mixture, wire showed strong anode aging.
EDX spectrum is taken from the place marked with the box.
    \label{gr53}}
\end{figure}

\begin{figure}
\caption{ SEM micrographs and EDX spectra for wires N2
and N5 irradiated at  2.25~kV (2.5~kV) up to radiation doses  
$\sim$~200~$\frac{mC}{cm \cdot wire}$ ($\sim$~800~$\frac{mC}{cm \cdot wire}$)
in an $Ar/CF_4/CH_4$ (67:30:3) mixture with
500~$ppm$ and 1400~$ppm$  $H_2O$.
  \label{ttt111}}
\end{figure}

\begin{figure}
\caption{ Typical deposits ('whiskers')
on the anode wires N5 (2.5~kV) and N9 (2.6~kV)
after irradiation in an $Ar/CF_4/CH_4$ (67:30:3) mixture.
Even in the 'whiskers' free region,
surface deposit is present, as can be evident from the EDX spectrum
taken in the place marked with box.
  \label{gr52}}
\end{figure}

\begin{figure}
\caption{ SEM micrographs 
of wire N1, irradiated in an 
$\afh$(67:30:3) mixture with 1400~$ppm$  $H_2 O$ and without water
up to an accumulated charge of 60~$\frac{mC}{cm \cdot wire}$,
show the different EDX sensitivity at  5~kV and 20~kV.
  \label{pic5}}
\end{figure}

\begin{figure}
\caption{ EDX spectroscopy of the cathodes in cells N5,N6,N8,N10. 
A thin deposit, consisting of  $C$, $F$ and $O$,
was detected on the irradiated $Al$ surfaces.
  \label{pic6}}
\end{figure}

\begin{figure}
\caption{ Efficiency profiles along wires N5,N10,N13,
measured in an electron beam at 2.35~kV,
after an accumulated charge of 700~$\frac{mC}{cm \cdot wire}$
in the  $Ar/CF_4/CO_2$ (65:30:5) + 1000~$ppm$ of $H_2O$ mixture.
These results are to be compared with efficiency of reference wires
$> 99~\%$ at 2.35~kV.
    \label{graf41}}
\end{figure}

\begin{figure}
\caption{ SEM imaging of the wires irradiated at
high voltages: 2.5~kV (wire N5), 2.6~kV (wire N10), 2.65~kV (wire N13)
up to a dose $\sim$~700~$\frac{mC}{cm \cdot wire}$.
Despite some local deterioration of the smooth wire surface
N13, the EDX spectrum at 20~kV is dominated by $Au$ peak.
  \label{pic11}}
\end{figure}

\begin{figure}
\caption{ EDX spectroscopy of the 'reference cell' cathode (N3)
and cathodes irradiated at 
high voltages: 2.5~kV (N5), 2.55~kV (N8), 2.65~kV (N13)
up to a dose $\sim$~700~$\frac{mC}{cm \cdot wire}$ in an
$Ar/CF_4/CO_2$ (65:30:5) + 1000~$ppm$ of $H_2O$ mixture.
  \label{pic22}}
\end{figure}


\newpage

\begin{thebibliography}{99}

\bibitem{Proposal} T. Lohse $et~al.$, HERA-B collaboration,
An Experiment to Study CP Violation in the B System Using an 
Internal Target at the HERA Proton Ring, Proposal,
{\bf DESY-PRC 94/04} (1994). 
 
\bibitem{TDR} E. Hartouni $et~al.$, HERA-B collaboration,
An Experiment to Study CP Violation in the B System Using an 
Internal Target at the HERA Proton Ring, Design Report,
{\bf DESY-PRC 95/01} (1995). 

\bibitem{sauli} F. Sauli, 
Nucl. Instr. and Meth. A{\bf 408} 258(1998).

\bibitem{schmidt} B. Schmidt {\it et al},
Nucl. Instr. and Meth. {\bf A 419} (1998) 230-238.

\bibitem{zeuner} T. Zeuner,  
Nucl. Instr. and Meth. {\bf A 446} (2000) 324-330.

\bibitem{stegmann} C. Stegmann,
Nucl. Instr. and Meth. {\bf A 453} (2000) 153-158.

\bibitem{hohlmann} M. Hohlmann,  
Nucl. Instr. and Meth. {\bf A 461} (2001) 21-24.

\bibitem{kolanoski}  H. Kolanoski, Investigation of Aging
in the HERA-B Outer Tracker Drift Tubes,
Proceedings of Nuclear Science Symposium and Medical Imaging 
Conference, 15-20 October 2000, Lyon, France. 

\bibitem{dissert1} A. Schreiner, Aging studies of drift chambers
of the HERA-B outer tracker using $CF_4$-based gases, 
Dissertation, Humboldt University, (2001).

\bibitem{max} M. Danilov, L. Laptin, I. Tichomirov, M.Titov, 
Yu. Zaitsev,  
Aging tests of proportional wire chambers using $Ar/CF_4/CH_4$
(74:20:6), $Ar/CF_4/CH_4$ (67:30:3) and $Ar/CF_4/CH_4$ (65:30:5) 
mixtures for the HERA-B Muon Detector, ITEP-43-00 (2000);
hep-ex/0107080.

\bibitem{algeri} A. Algeri {\it et al},
Nucl. Instr. and Meth. {\bf A 338} (1994) 348-367.

\bibitem{tsarnas} N. Spielberg, D. Tsarnas
Rev. Sci. Instr. {\bf Vol.46 (8)} (1975) 1086-1091.

\bibitem{fraga} M. Fraga {\it et al},
Nucl. Instr. and Meth. {\bf A 419} (1998) 485-489.

\bibitem{max1} 
The effect of added $CF_4$ on aging performance of wire
proportional chambers, in preparation.

\bibitem{electronics} M. Buchler {\it et al},
IEEE Trans. Nucl.Sci., {\bf NS-46 } (1999) 126-132.

\bibitem{max2}  S. Barsuk {\it et al}, 
A Gaseous Muon Detector at the HERA-B Experiment,
IEEE Trans. Nucl.Sci., {\bf NS-48(4)} (2001) 1059-1064.

\bibitem{kushner} M.J. Kushner,
J. Appl. Phys, {\bf Vol.53, (4)} (1982) 2923-2938.

\bibitem{chemmod} J. Wise {\it et al}, 
J. Appl. Phys, {\bf Vol.74, (9)} (1993) 5327-5340.

\bibitem{dissert} V. Pashhoff, 
Dissertation, University Freiburg, October (1999)

\bibitem{kolly1} M. Kollefrath, 
Dissertation (in german), University Freiburg, (1999)

\bibitem{denisov} D.S. Denisov, 
On using $CF_4$ as a working gas for drift tubes (in russian),
IHEP-preprint-90-16 (1990)

\bibitem{kay} E. Kay, Invited Pap. Int. Round Table Plasma Polym.
Treat., IUPAC Symp., Plasma Chem. (1977).

\bibitem{yasuda} H. Yasuda, Plasma Polymerization,
(Academic Press, 1985)

\bibitem{mogab} C. Mogab {\it et al}, 
J. Appl. Phys, {\bf Vol.49, (7)} (1978) 3796-3803.


\bibitem{opensh} R. Openshaw {\it et al},
Nucl. Instr. and Meth. {\bf A 307} (1991) 298-308.

\bibitem{d0} G. Alexeev {\it et al},
Technical design Report for the D0 Forward Muon Tracking
Detector Based on Mini-Drift Tubes, D0 Note 3366, (1997)

\bibitem{fast1} L.G. Christophorou {\it et al}, 
Nucl. Instr. and Meth. {\bf A 163} (1979) 141-149.

\bibitem{fischer} J. Fischer {\it et al}, 
Nucl. Instr. and Meth. {\bf A 238} (1979) 249-264.

\bibitem{schmidt1} B. Schmidt, S. Polenz 
Nucl. Instr. and Meth. {\bf A 273} (1988) 488-493.

\bibitem{cf41} L.G. Christophorou {\it et al}, 
J. Phys. Chem. Ref. Data, {\bf Vol.25, No.5} (1996) 1341-1388.

\bibitem{kadyk1} J. Kadyk {\it et al},
IEEE Trans. Nucl. Sci {\bf NS-37 (2) } (1990) 478-486.

\bibitem{opensh1} R. Openshaw {\it et al},
Nucl. Instr. and Meth. {\bf A 307} (1991) 298-308.

\bibitem{workshop} Proc. Workshop on Radiation Damage to Wire Chambers, 
Lawrence Berkeley Laboratory (Jan. 1986) LBL-21170.

\bibitem{vavra} J. Va'vra, ref.~\cite{workshop}, pp. 263-294.

\bibitem{kadyk} J. A. Kadyk,
Nucl. Instr. and Meth. {\bf A 300} (1991) 436-479.

\bibitem{arikado} T. Arikado, Ya. Horiike,
Jpn. J. Appl. Phys, {\bf Vol.22, (5)} (1983) 799-802.

\bibitem{martz} J.C. Martz {\it et al},
J. Appl. Phys, {\bf Vol.67, (8)} (1990) 3609-3617.

\bibitem{gasflow} E. Truesdale, G. Smolinsky,
J. Appl. Phys, {\bf Vol.50, (11)} (1979) 6594-6599.

\bibitem{gasflow1} E. Truesdale {\it et al},
J. Appl. Phys, {\bf Vol.51, (5)} (1979) 2909-2913.


\bibitem{romaniouk} A. Romaniouk,
 Choice of materials for the constructio of TRT,
ATLAS Internal Note, INDET-98-211 (1998)


 
\end{thebibliography}
\end{document}